\documentclass[twocolumn,aps,pra,showpacs,preprintnumbers,amsmath,amssymb]{revtex4}
\usepackage{natbib}
\usepackage{graphicx}
\usepackage{bm}
\usepackage{color}
\begin{document}
\title{Decoherence dynamics of interacting qubits coupled to a bath of local optical phonons}
\author{ Muzaffar Qadir Lone\footnote{email: lone.muzzafar@gmail.com}}
\affiliation{
Department of Physics, University of Kashmir, Srinagar-190006,India}
\affiliation{
TCMP Division,
1/AF Salt Lake, Saha Institute of Nuclear physics, Kolkata, India.}
\author{ S. Yarlagadda\footnote{email: y.sudhakar@saha.ac.in}}
\affiliation{
TCMP Division,
1/AF Salt Lake, Saha Institute of Nuclear physics, Kolkata, India.}

\date{\today}

\begin{abstract} 
We study decoherence  in an 
interacting qubit system described by infinite range Heisenberg model (IRHM) 
in a  situation where the system is coupled to a bath of local optical phonons. 
 Using perturbation theory in polaron frame of reference, we derive an effective Hamiltonian that is valid in the regime 
of  strong  spin-phonon coupling under non-adiabatic conditions. 
It is shown that the effective Hamiltonian commutes with the IRHM upto leading orders of perturbation
 and thus has the same eigenstates as the IRHM.
 Using a quantum master equation with Markovian approximation of dynamical evolution, 
we show that the off diagonal elements of the density matrix donot decay in the energy eigen basis 
of IRHM.
\end{abstract}
\maketitle

\section{Introduction} 

A closed isolated quantum system will always follow unitary quantum dynamics given by Schrodinger
equation. However every  quantum system that we try to study or model is inevitably coupled to some 
form of environment and hence an open quantum system \cite{Pet,Uweiss}. The coupling of a 
quantum system with its environment  leads to decoherence, the process
 by which information is degraded.
Decoherence is  the fundamental mechanism by which fragile superpositions are destroyed
thereby producing a quantum to classical transition \cite{schloss,zurek2}. 
In fact, decoherence is one of the main obstacles for the preparation, observation, and
implementation of multi-qubit entangled states.
The intensive work on quantum information and computing in recent years has 
tremendously increased the
interest in exploring and controlling decoherence effects   
\cite{nat1,milb2,QA,diehl,verst,weimer,bellmo,FR1,FR2,FR3,FR4,FR5,FR6,FR7,FR8,FR9,FR10,FR11,FR12}.

 The dynamics of an open quantum  system coupled to a bath can be either Markovian or non-Markovian 
\cite{Pet,RLF,BBell,BBell1,BBell2,AR,SM}.
 However in this paper we are concerned with the Markovian dyanmics of the system infinite range Heisenberg model (IRHM) coupled to
 a bath of local optical phonons.
 In case of Markovian processes, the environment acts as a sink for the system information; the 
system of interest loses information into the environment and this lost information plays no role in 
the dynamics of the system \cite{Pet,NC}. Although the theory of decoherence has undergone major 
advances \cite{schloss,zurek2}, 
yet, there exist many definitions of decoherence \cite{coles}.
For the analysis in this paper, we choose the most commonly used
definition of decoherence:  Loss of off-diagonal elements
in the system's reduced density matrix.
In general, a many-qubit (i.e., many-spin) system can have 
distance dependent interaction. The two limiting cases for interaction are
spin interactions that are independent of distance and spin chains with nearest-neighbor
interactions only. In this work we consider the extreme case of distance independent interaction
among the spins, i.e., the IRHM.

In this paper,  we employ 
the analytically simpler frame of reference
of hard-core-bosons (HCBs) rather than that of spins so that the single particle excitation spectrum 
can be easily 
obtained and exploited; we show that the effective
Hamiltonian 
even in higher order (i.e., greater than second order)  perturbation theory
retains the same eigenstates as
the IRHM when the spins are coupled to local phonons. Furthermore, decoherence is studied using
 the quantum master equation approach {\cite{nazir}}. 
 Our analysis based on Markovian quantum master equation shows that the
 off diagonal matrix elements of density matrix in eigen-basis of IRHM do not decay.

The rest of the paper is organized as follows:
In section II, we introduce the IRHM Hamiltonian and map the Hamiltonian to a HCB model. Also in 
the same section we transform IRHM to a polaron frame using a canonical 
transformation. 
In section III,
we use second order perturbation theory and with the help of Schrieffer-Wolf (SW) transformation, 
we derive an effective Hamiltonian that commutes with $H_{\rm IRHM}$ and thus have same set of 
eigenstates. In section IV, 
we use the master equation approach 
and show that the system does not decohere under 
Markovian approximation.
 Finally we conclude in section V  and make some 
general remarks
regarding the wider context of our results. 
The paper also contains  an  Appendix A, where  we derive the third order perturbation
contribution to
our effective Hamiltonian ($H_{eff}$) and show that the eigenstates of the IRHM Hamiltonian are 
retained
by our $H_{eff}$.

\section{Model Hamiltonian for the system coupled to a bath of phonons }

We consider a system of spin-$\frac{1}{2}$ particles interacting with each other  through 
 a infinite range  anisotropic Heisenberg antiferromagnetic exchange interaction i.e. IRHM:

\begin{eqnarray}
 H_{\rm IRHM} = J \sum_{i,j>i} \!\! \left [ \vec{S_i}.\vec{S_j} + (\Delta 
-1)S^z_i S^z_j \right]
\label{Hs}
\end{eqnarray} 
where $J > 0$, $\Delta \geq 0$, and  $S_i=\frac{1}{2} \sigma_i$, $i=x,~y,~z$.
We note that $H_{\rm IRHM}$
 commutes with both  $S^z_{Total}$ ($\equiv \sum_i S^z_i$) and $\left ( \sum_{i} \vec{S_i} \right 
)^2$ ($\equiv S^2_{Total}$).
In equation (\ref{Hs}), it is understood that $J = J^{\star}/(N-1)$ 
(where $ J^{\star}$ is a finite
quantity) so that the energy per site remains finite as $N \rightarrow \infty$.
The eigenstates of $H_{\rm IRHM}$ are characterized by $S_T$ (i.e., the total spin eigenvalue)
and $S^z_T$ (or the eigenvalue of the z-component of the total spin $S^z_{Total}$).
 The ground state corresponds to $S^z_T =0$ and $S_T=0$ which is SU(2)
invariant.

The IRHM has relevance to many physical problems. The Lipkin-Meshkov-Glick (LMG) model \cite{LMG} 
$H_{\rm LMG} = -2h(\sum_jS_j^z) -2\lambda[(\sum_j S_j^x)^2+\gamma (\sum_jS_j^y)^2]/N$͒
well studied in nuclear many body problem (for $h=0$ and $\gamma=1$) is a special case of the above mentioned 
long-range model for certian set of paramters. 
It has been shown by Ezawa that the long-range ferromagnetic Heisenberg model describes
well a zigzag graphene nanodisc \cite{ezawa}.  For  spin systems with spins defined on the corners of a regular tetrahedron can be realized
(from a Hubbard model) as exact special cases of the above  long-range model \cite{Hubbard}.
In solid state quantum computation using semiconductor quantum dots, 
spin states are prepared, manipulated, and measured using rapid control of Heisenberg exchange interaction \cite{semidqd}.

The real quantum computer will not be free from noise
and thus the entangled states have a tendency to undergo decoherence.
To study decoherence due to phonons, we consider interaction with
 optical phonons such as would
be encountered when considering transition metal oxides.
We write the  total Hamiltonian $H_T$ as 
\begin{eqnarray}
\!\!\!\! H_T = H_{\rm IRHM}+ g \omega\sum_i S^z_i (a^{\dagger}_i + a_i) + \omega \sum_i 
a^{\dagger}_i a_i ,
\label{Ham3}
\end{eqnarray}
where $a$ is the phonon destruction operator \cite{indexk}, $\omega$ is the optical phonon 
frequency, and $g$ is the
coupling strength. In order to make the calculations simple
from spin excitations to particle excitations in our model, we  make the mapping of spin operators for spin-$\frac{1}{2}$
 particles on HCBs. 
HCBs are defined on lattice sites $i = 1, ..., N$
with restricted occupation numbers, $n_i= 0, 1$ \cite{AssA}. The constrained
creation and destruction operators $b^{\dagger}$ and $b$,  are defined as  
$b^{\dagger} = S^{+}$, $b = S^{-}$, and $b^{\dagger} b = S^{z} + 0.5$. We then
observe that conservation of $S^z_{Total}$ implies conservation of total number of HCB. 
The total Hamiltonian is then given by
\begin{eqnarray}
H &=&  J \sum_{i,j>i}[(0.5 b^{\dagger}_i  b_{j} +{\rm H.c.}) + \Delta (n_i - 0.5)(n_j -0.5)  ] 
\nonumber \\
          &&+ \omega \sum_j a^{\dagger}_{j} a_j
          + g \omega \sum_j (n_j-\frac{1}{2}) 
 (a_j +a^{\dagger}_j) , 
\end{eqnarray}
where
 $n_j \equiv b^{\dagger}_{j} b_j $.
Subsequently, we perform 
the well-known Lang-Firsov (LF) transformation \cite {lang,sdadys} on this
Hamiltonian. 
Under the LF transformation  given by $e^S H e^{-S} =H_0+H_I $ with
$S= - g \sum_i (n_i-\frac{1}{2}) (a_i - a^{\dagger}_i)$, the operators
$b_j$ and $a_j$ transform like fermions  and bosons;
this is due to the interesting 
 commutation properties of HCB given below: 
\begin{eqnarray}
[b_i,b_j]&=&[b_i,b^{\dagger}_j]= 0 , \textrm{ for } i \neq j , \nonumber\\
\{b_i,b^{\dagger}_i\}& = & 1 .
\label{commute}
\end{eqnarray}   
Next, the unperturbed Hamiltonian $H_0$ is expressed as 
\cite{sdadys}
\begin{eqnarray}
\!\!
H_0 =  H_s +H_{env} ,
\label{H0}
\end{eqnarray}
where  we identify $H_s$ as the system Hamiltonian
\begin{eqnarray}
\!\!\!\! H_s &=& J \sum_{i,j>i}[(0.5 e^{-g^2}b^{\dagger}_i  b_{j} +{\rm H.c.})
\nonumber \\
&& ~~~~~~~~~~ + \Delta (n_i -0.5)( n_j - 0.5)  ] ,
\label{HHs}
\end{eqnarray}
and $H_{env}$ as the Hamiltonian of the environment
\begin{eqnarray}
H_{env} = \omega \sum_j a^{\dagger}_j a_j . 
\end{eqnarray}
On the other hand, the interaction $H_I$ which we will treat as perturbation is given by
\begin{eqnarray}
H_I
= J \sum_{i,j>i}[0.5 e^{-g^2}b^{\dagger}_i  b_{j} ]
            \{\mathcal S^{{ij}^\dagger}_+ \mathcal S^{ij}_{-}-1\} +{\rm H.c.} ,
\label{int}
\end{eqnarray}
where $\mathcal S^{ij}_{\pm} = \textrm{exp}[\pm g(a_i - a_{j})]$.
 { In the transformed frame, the system Hamiltonian depicts that all the HCBs are coupled to the 
same
phononic mean-field. Thus, the unperturbed Hamiltonian $H_0$ comprises of
the system Hamiltonian $H_s$ representing
  HCBs with the reduced hopping term $0.5 J e^{-g^2}$
and the environment Hamiltonian $H_{env}$ involving displaced bath oscillators corresponding to 
local distortions.
Here it should be pointed out that both the 
interaction of the HCB with the mean-field as well as the local polaronic distortions in the bath 
oscillators
  involve controlled degrees of freedom. { Now, the system Hamiltonian $H_s$ can be expressed as} 
  \begin{equation}
   H_s=H_{\rm IRHM}+(H_s-H_{\rm IRHM})
  \end{equation}
  {When we change the Hamiltonian from  $H_{\rm IRHM}$ to $H_s$ by adiabatically turning on the 
perturbation $(H_s-H_{\rm IRHM})$,}
the resulting state of the system is still obtainable from that of $H_{\rm IRHM}$ by using unitary
Hamiltonian dynamics 
  and is thus predictable based on a knowledge of the coupling parameter $g$ \cite{gl}.
Thus  no irreversibility
is involved in going from $H_{\rm IRHM}$ to $H_s$.
 {\it On the other hand, perturbation $H_{I}$ pertains to the interaction of HCBs with local 
deviations from the phononic mean-field;
  the interaction term $H_I$ represents numerous
or uncontrolled environmental degrees of freedom and thus has the potential
for producing decoherence}. Furthermore, it is of interest to note that
the interaction term is weak in the transformed frame
compared to the interaction 
in the original frame; thus one can perform perturbation theory with the interaction term.
}
 
\section{Effective Hamiltonian from second  order perturbation theory}

In this section we derive an effective Hamiltonian using second order perturbation theory and Schrieffer-Wolff
 (SW) transformation.
 We represent the  eigenstates of the unperturbed Hamiltonian $H_{0}$ as
$|n,m\rangle\equiv|n\rangle_{s}\otimes|m\rangle_{ph}$ with the corresponding
 eigenenergies $E_{n,m}=E_{n}^{s}+E_{m}^{ph}$;
$|n\rangle_{s}$ is the eigenstate of the system with eigenenergy $E_{n}^{s}$
while $|m\rangle_{ph}$ is the eigenstate for  the environment with eigenenergy $E_{m}^{ph}$.
 Henceforth, for brevity, we will use $\omega_m \equiv E_m^{ph}$.
On observing that $\langle0,0|H_{I}|0,0\rangle=0$ ({ i.e., the ground state
expectation value of the deviations is zero}),
we obtain the next relevant second-order perturbation term \cite{sdadys}
\begin{eqnarray}
E^{(2)}=\sum_{n,m}{{\langle0,0|H_{I}|n,m\rangle\langle n,m|H_{I}|0,0\rangle}\over{E_{0,0}-E_{n,m}}}.
\end{eqnarray}

Employing the SW transformation (see Appendix A of Ref \cite{srsypbl}) with the conditions of  strong coupling ($g>1$) 
and non-adiabaticity  ($J^{\star}/\omega \leq 1$), we 
get the following second-order term $H^{(2)}$ \cite{sdys}
\begin{eqnarray}
\!\!\!\!H^{(2)} \! &=& \!
-\sum_{m}{{{_{ph}\langle0|H_{I}|m\rangle_{ph}}~{_{ph}\langle 
m|H_{I}|0\rangle_{ph}}}\over{\omega_{m}}} 
\nonumber \\
&=& \! \sum_{i, j > i } 
 \left [(0.5 J_{\perp}^{(2)}
b^{\dagger}_i
 b_j +{\rm H.c.}) \right .
\nonumber \\
&&~~~~~ \left . - 
 0.5 J_{\parallel}^{(2)}
 \{n_i(1-n_j)+n_j(1-n_i)\} \right ] ,
\label{H_eff}
\end{eqnarray}
where 
\begin{eqnarray}
\!\!\!\!\!\!\!\!\!\!\!\! J_{\perp}^{(2)} \equiv  - (N-2) f_1 (g) \frac{J^2 e^{-2g^2}}{2 \omega}
 \sim -(N-2) \frac{J^2e^{-g^2}}{2g^2 \omega} ,
 \end{eqnarray}
\begin{eqnarray}
 J_{\parallel}^{(2)} \equiv 
[2f_1 (g)+f_2(g)]\frac{J^2 e^{-2g^2}}{{2 \omega}} 
\sim \frac{J^2 }{{4g^2 \omega}}, 
 \end{eqnarray}
 with
$f_1(g) \equiv \sum^{\infty}_{n=1} g^{2n}/(n!n)$
 and
$f_2(g) \equiv \sum^{\infty}_{n=1}\sum^{\infty}_{m=1} g^{2(n+m)}/[n!m!(n+m)]$. 
{The effective Hamiltonian $H_{s}+H^{(2)}$ is a low energy Hamiltonian
obtained by the canonical SW transformation \cite{schrieffer,loss2} decoupling the low-energy and 
the
high-energy subspaces; this decoupling  is a consequence of $J^{\star}e^{-g^2} \ll \omega$. }
We now make the important observation that 
the effective Hamiltonian $H_{s}+H^{(2)}$, when expressed in terms of spins, 
 has the following form:
\begin{eqnarray}
 \sum_{i, j > i } 
 \left [ J_{\rm tr} ({S_i^{x}}{S_j^{x}} + {S_i^{y}}{S_j^{y}}) + J_{\rm lng} S_i^zS_j^{z} \right ] ,
\label{Heff}
\end{eqnarray}
and thus has eigenstates identical to
those of the original Hamiltonian $H_{\rm IRHM}$ in equation (\ref{Hs})
because $\sum_{i, j > i } 
 (S_i^{z} S_j^{z})$ and $ H_{\rm s}$ commute. 
On carrying out higher order 
(i.e., beyond second order) perturbation theory (as discussed in Appendix A),
and expressing the results in the spin language,
we still get an effective Hamiltonian $H_{eff}$ of the following form that
has the same eigenstates as the s.
\begin{eqnarray}
H_{eff} \! &=& \! \sum_{i, j > i } 
 \left [ J_{xy}  ( \sum_k S_k^z ) ({S_i^{x}}{S_j^{x}} + {S_i^{y}}{S_j^{y}}) \right ]
\nonumber \\
 &&+
 \sum_i J_{z}  (\sum_k S_k^z  )S_i^{z} ,
\label{Heff}
\end{eqnarray} 
where $J_{xy}$ and $J_z$ are functions of the $S^z_{Total}$  ($= \sum_k S_k^z$ ) operator.
It is the infinite range of the Heisenberg model that enables  
 the eigenstates of the system to 
remain unchanged.
Next, we study decoherence in a dynamical
context and gain more insight into how the states
of our $H_{\rm s}$ can be decoherence free.

{\section{Markovian Dynamics}}
In this section, we study the markovian dynamics of our system in polaron frame of reference.
The dynamics of the system, described by the reduced density matrix $\rho_s(t)$ at 
time $t$, is obtained from  the density matrix $\rho_T(t)$ of the total system by taking the 
partial trace over the degrees of freedom of the environment:
\begin{eqnarray}
\rho_s(t) = Tr_R\left[\rho_T(t)  \right]= Tr_R\left[  U(t) \rho_T(0) U^{\dagger}(t)     \right] ,
\end{eqnarray}
where $U(t)$  represents the time-evolution operator of the total system. Now it is
evident from the above equation that we need first to determine 
the dynamics of the total system which is a  difficult task 
in most of the cases. 
By contrast, master equation approach 
conveniently  and directly yields the time evolution of
the reduced density matrix 
 of the system interacting  with an environment.
This approach relieves us from the need of having to first determine the dynamics of the
total system-environment combination and then to trace out the degrees
of freedom of the environment.

We consider  the following Hamiltonian:
\begin{eqnarray}
H=  H_0 + H_I ,
\end{eqnarray}
where $H_0$ is the system-environment Hamiltonian given by equation (\ref{H0}) 
and $H_I$ represents the interaction Hamiltonian given by equation (\ref{int}).
Defining an operator $O$ in interaction picture as $\tilde{O}= e^{i H_0 t} O e^{- i H_0 
t}$,  we write the quantum master equation in Born approximation \cite{Pet}

\begin{eqnarray}
\!\!\!\!\!\!\!\!\!\!
\frac{d \tilde{\rho}_s(t)}{dt} &=& 
-i~Tr_R[\tilde{H}_I(t), \rho_s(0) \otimes R_o] \nonumber \\
&-& \int_0^t d\tau Tr_R[\tilde{H}_I(t),[\tilde{H}_I(t-\tau), \tilde{\rho}_s(t) \otimes R_0]] 
\label{mas3}
\end{eqnarray}
where $R_0=\sum_{n}|n \rangle_{ph}~\!\! _{ph}\langle n|e^{-\beta \omega_n}/Z$ is the bath density 
matrix with $Z$ as the partition function.
In order to  study the Markovian dynamics of the system,  we assume that 
the correlation time scale  $\tau_{c}$  for the environmental fluctuations 
is negligibly small compared to the relaxation time scale $\tau_{s}$  for the system,
i.e.,  $\tau_{c} \ll \tau_{s}$.
The time scale over which the system changes is $\tau_s \sim \frac{1}{J^{\star}e^{-g^2}}$
and the bath correlation time scale is $\tau_c \sim \frac{1}{\omega}$. The Markovian approximation is motivated 
by the condition 
 $J^{\star}e^{-g^2} \ll \omega$ already mentioned in section III. 
The Markov approximation ($\tau_{c} \ll \tau_{s}$)
allows us to set the upper limit of the integral to $\infty$ in equation (\ref{mas3}).
 Thus we obtain the second order time-convolutionless
 Markovian quantum master equation 
\begin{eqnarray}
\!\!\!\!\!\!\!\!\!\!
\frac{d \tilde{\rho}_s(t)}{dt} &=& 
-i~Tr_R[\tilde{H}_I(t), \rho_s(0) \otimes R_o] \nonumber \\
&-& \int_0^\infty d\tau Tr_R[\tilde{H}_I(t),[\tilde{H}_I(t - \tau), \tilde{\rho}_s(t) \otimes R_0]].
\label{mark}
\end{eqnarray}
Defining $\{|n\rangle_{ph}\}$ as the basis set for phonons, therefore, we can write
the master equation as (See Appendix B for details):
\begin{widetext}

\begin{eqnarray}
~~~~~~~~~~~ \frac{d \tilde{\rho}_s(t)}{dt} 
&=& -\sum_{m} \int_0^\infty d\tau \left[ |_{ph}\langle 0 | H_I |m\rangle_{ph}|^2
 ~\tilde{\rho}_s(t)  e^{-i\omega_m\tau}  + \tilde{\rho}_s(t) ~ |_{ph}\langle 0 | H_I 
|m\rangle_{ph}|^2 
  e^{i\omega_m\tau} \right] \nonumber \\
 &&+ \sum_{n} \int_{0}^\infty d\tau \left[ _{ph}\langle n | H_I |0\rangle_{ph}  \tilde{\rho}_s(t) 
_{ph}\langle 0|
 H_I|n\rangle_{ph} e^{i\omega_n\tau} \right .
\nonumber \\
&& ~~~~~~~~~~~~~~~~~~ \left . +  _{ph}\langle n | H_I |0\rangle_{ph}  \tilde{\rho}_s(t) _{ph}\langle 
0| H_I|n\rangle_{ph}
 e^{-i\omega_n\tau} \right]  \nonumber \\
&=& - \sum_n  
\left[ \int_0^{\infty} d\tau ~ e^{- i(\omega_n -i \eta)\tau} |_{ph}\langle 0 |H_I |n \rangle_{ph}|^2 
~ \tilde{\rho}_s(t)
\right . 
\nonumber \\
&& ~~~~~~~~~~ + \left .
\int_{0}^{\infty} d\tau ~ e^{i(\omega_n+i\eta)\tau}
 ~\tilde{\rho}_s(t) ~|_{ph}\langle 0 |H_I |n \rangle_{ph}|^2  \right .\nonumber \\
&&~~~~~~~~~~ - \left. \int_{-\infty}^{\infty} d\tau ~ e^{i\omega_n \tau} ~ _{ph}\langle n| H_I 
|0\rangle_{ph} 
~\tilde{\rho}_s(t)~ _{ph}\langle 0|H_I|n\rangle_{ph} 
    \right]. \nonumber \\
\label{21}
\end{eqnarray}

\end{widetext}

Now, we know $\int_{-\infty}^{\infty} d\tau e^{i \omega_n \tau} \propto \delta(\omega_n)$.
Therefore, on using this relation and  the fact that ${_{ph}\langle 0| H_I |0\rangle_{ph}} =0$, the 
third term in equation (\ref{21}) vanishes; hence, we get
\begin{eqnarray}
\!\!\!\!\!\!
\frac{d \tilde{\rho}_s(t)}{dt} =  i\sum_n \left[
  \frac{ |_{ph}\langle 0 |H_I |n \rangle_{ph}|^2 }{\omega_n} \tilde{\rho}_s(t) - 
\tilde{\rho}_s(t) \frac{ |_{ph}\langle 0 |H_I |n \rangle_{ph}|^2 }{\omega_n} 
\right] \nonumber \\ .
\end{eqnarray}
The term $\sum_n \left[\frac{ |_{ph}\langle 0 |H_I |n \rangle_{ph}|^2 }{\omega_n} \right]$ 
corresponds to the effective Hamiltonian  $H^{(2)}$  in second order perturbation and commutes
 with $H_0$ (see section III). Let $|n\rangle_s$ be the simultaneous eigenstate for $H^{(2)}$ and 
$H_s$ with
 eigenvalues $E_n^{(2)}$ and $E_n^s$, respectively. Then, from the above equation we get:
\begin{eqnarray}
{_s\langle n|\tilde{\rho}_s(t) | m \rangle_s} &=&
 e^{-i(E^{(2)}_n- E_m^{(2)})t}~ _s\langle n|\tilde{\rho}_s(0) | m \rangle_s ,
\end{eqnarray}
which implies that
\begin{eqnarray}
 {_s\langle n|\rho_s(t) | m \rangle_s}
& =& e^{-i(E_n- E_m)t}~ _s\langle n|\rho_s(0) | m \rangle_s ,
\label{sol}
\end{eqnarray}
where $E_n= E_n^s+E_n^{(2)}$.
Thus we see from the above equation that there is only a phase shift but no decoherence!
{Since the matrix elements of an operator are invariant
under canonical transformation, thus under Markovain dynamics, it  should be clear that no loss in off-diagonal density matrix 
elements 
(i.e., no decoherence)
 in the LF transformed 
frame of reference implies no loss in off-diagonal density matrix elements (i.e., no decoherence)
in the
original untransformed frame of reference.
Although the HCB's in the original frame of reference form polarons and are thus
entangled with the environment, nevertheless no decoherence results.
For greater clarity, we take the example of two qubit state of IRHM i.e N=2. 
From  equation (\ref{sol}), the matrix element $_s\langle n|\rho_s(t) | m \rangle_s$ can be written as
{\begin{eqnarray}
 _s\langle n|\rho_s(t) | m \rangle_s  
&=& _s\langle n|~\Bigg[\sum_{n}~ _{ph}\langle n| \rho_T(t) | n \rangle_{ph}\Bigg]~| m \rangle_s 
\nonumber \\
&=&~  _s\langle n|\sum_{n}~_{ph}\langle n| e^S \rho^o_T(t)
    e^{-S} | n \rangle_{ph}~| m \rangle_s ,
 \nonumber \\
\label{den}
\end{eqnarray}}
where $\rho^{o}_T(t)$ is the total density matrix in the original frame of reference .
Now, we illustrate this quantity by considering the simple two-spin (i.e., N=2) case of the IRHM.
The singlet state $\frac{1}{\sqrt{2}}(| \uparrow \downarrow \rangle - | \downarrow \uparrow 
\rangle)$
 and the triplet state  
$\frac{1}{\sqrt{2}}(| \uparrow \downarrow \rangle + | \downarrow \uparrow \rangle)$
are the  $S^z_T=0$ eigenstates
of the two-qubit IRHM Hamiltonian; in HCB language, these states are expressed as
$\frac{1}{\sqrt{2}}(| 10 \rangle - | 01 \rangle)$ 
and $\frac{1}{\sqrt{2}}(| 10 \rangle + | 01 \rangle)$, respectively.
Now, the operator $e^{-S}$ can be expressed as
{\begin{eqnarray}
 e^{-S} &=& e^{g\sum_{i=1,2}(n_i-\frac{1}{2}) (a_i - a^\dagger _i)} \nonumber \\
&=& \prod_{i=1,2} e^{g(n_i-\frac{1}{2})(a_i - a^\dagger _i)} \nonumber \\
&=& \prod_{i=1,2} \Bigg[n_iX_i + (1-n_i)X_i^{\dagger}\Bigg] ,
\end{eqnarray}}
where $X_i=e^{\frac{g}{2}(a_i - a^\dagger _i)}$.
Using the above, we obtain
\begin{eqnarray}
 &&e^{-S}\frac{1}{\sqrt{2}}(| 10 \rangle \pm | 01 \rangle) | m_1, m_2 \rangle_{ph} \nonumber \\
&&~~~~~~~~~~~~
= [{X}_1 {X}^{\dagger}_2 |10\rangle \pm {X}_2 
{X}^{\dagger}_1  |01\rangle ]| m_1, m_2 \rangle_{ph} 
.
\end{eqnarray}
{where $m_1$ and $m_2$ correspond to phonon occupation numbers at site $1$ and site $2$ 
respectively.}
Therefore, from equation (\ref{den}) we can write the density matrix element between singlet and 
triplet states in the original frame of reference as
\begin{widetext}
{\begin{eqnarray}
\frac{1}{2} \left ( \langle10 | -  \langle01 | \right )&& \rho_{s}(t) 
\left ( | 10 \rangle + | 01 \rangle \right )   \nonumber \\
 && =\frac{1}{2}\sum_{m_1,m_2} {_{ph}}\langle m_1,m_2 | 
\left ( \langle10 | {X}_{2}{X}^{\dagger}_1 - 
 \langle01 | {X}_{1}{X}^{\dagger}_2 \right )
  \rho^o_{T}(t) 
  \left ( {X}_1 {X}^{\dagger}_2| 10 \rangle
 + {X}_2 {X}^{\dagger}_1| 01 \rangle \right ) | m_1,m_2 \rangle_{ph} .
 \nonumber \\
\label{off-dia}
\end{eqnarray}}
\end{widetext}
Depending upon the presence or absence of HCB, appropriate
deformation will be produced at each site and
$\left [\left ({X}_1{X}^{\dagger}_2| 10 \rangle \pm
{X}_2 {X}^{\dagger}_1| 01 \rangle \right )| m_1,m_2 \rangle_{ph} \right ]$ represents polaronic 
states.
Furthermore, in equation (\ref{off-dia}),
no loss in the off-diagonal matrix element on the left hand side 
implies no loss in the off-diagonal matrix element on the right hand side (i.e., 
no loss in the
measured density matrix elements in the original frame of reference) which in turn means no 
decoherence results.

 Thus, up to second order in perturbation, the 
 assumption $J^{\star} e^{-g^2} < < \omega$, the infinite range
of the Heisenberg model, and the Markov approximation 
 ($\tau_{c} \ll \tau_{s}$) together have ensured that the 
system, with a fixed  $S_T^{z}$, does not decohere.}
While the above analysis is valid in the regime $k_B T /\omega << 1$, the finite temperature
case  $k_B T /\omega \gtrsim 1$ needs additional extensive considerations
and will be dealt seperately.\\

\section{  Conclusions}
  In conclusion, we have shown that the eigenstates of $H_{eff}$ are the same as those of $H_{\rm 
IRHM}$ upto the leading order of perturbation. Also we have shown that 
the off-diagonal elements of the reduced density matrix donot deacy in polaron frame of reference, thus no decoherence results.
 {More specifically,  for local phonons, $_s\langle n|\rho_s(t) | m \rangle_s$ 
differs from $_s\langle n|\rho_s(0) | m \rangle_s$ only by a phase factor  
  and  $_s\langle n|\rho_s(0) | m \rangle_s$ can be obtained from 
{$_s\langle n|\rho_{\rm{IRHM}} | m \rangle_s$}
 (density matrix element of IRHM) by an exact unitary evolution \cite{gl}.}
  It would be of cosiderable interest to analyze the non-Markovain decoherence dynamics in the 
system and is left as future exercise.

Next, our decoherence analysis for local optical phonons will continue to be valid
 even for the more general optical phonon terms given below:
\begin{eqnarray}
\frac{1}{N^{1/2}}\sum_{i,k} S^z_i 
 [ \omega_k (g_k a^{\dagger}_{k,i} + g_k^{\star}a_{k,i})] 
 +  \sum_{k,i} \omega_k a^{\dagger}_{k,i} a_{k,i} .
\end{eqnarray}
We also must mention that our approach cannot accommodate the acoustic phonon case as here 
the condition $J^{\star}e^{-g^2} << \omega_k$
 cannot be satisfied in the long wavelength limit.

Next, we make a remark on applicability of our model and the decoherence in some real processes.
Understanding the highly efficient
transport of absorbed light-energy through molecules in
photosynthesis is of significant scientific interest and also
key to designing light-harvesting technology \cite{engel,Fleming,Alex1}.
The model that is   used for the study of the  excitation energy in 
Fenna-Matthews-Olson (FMO) complexes is an extreme long range interaction model \cite{Alex1} for excitons with uniform hopping strength between any pair of 
chromophores in FMO  complexes. The phonon fluctuations at various chromophores are uncorrelated to each other \cite{Fleming} i.e., 
local phonon effects are significant in such complexes. 
The system-bath coupling in photosynthetic complexes is
thought to be not weak but to be at least in the intermediate regime \cite{Fleming}; instead of employing the usual quantum
master equation techniques valid for the weak-coupling
limit, LF transformed master equation can be used.

\appendix

\section{}

\begin{figure}[]
\begin{center}
\includegraphics[width=3.5in,height=3.0in]{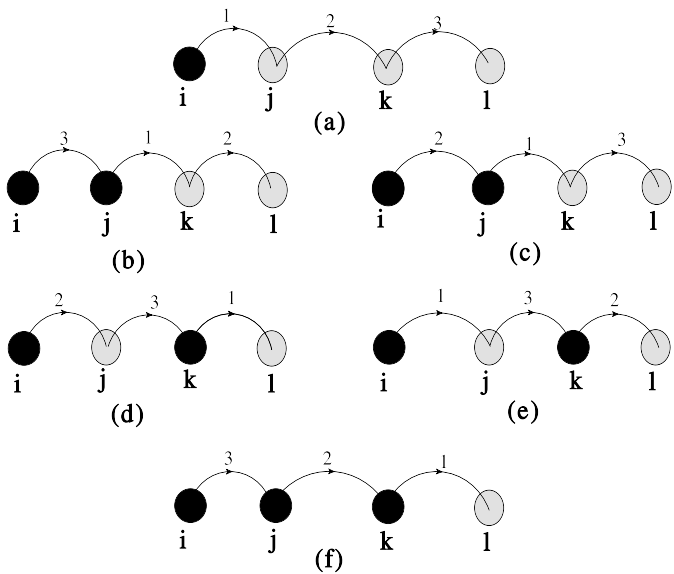}
\caption{ Open loop hopping processes contributing to effective hopping term $T_n^{li}$ in
third-order 
perturbation theory.
Here empty circles
correspond to sites with no particles while filled circles correspond 
to sites with hard-core-bosons. The numbers 1, 2, and 3 indicate the order of
hopping.
}
\label{fey1}
\end{center}
\end{figure}
\begin{figure}[]
\begin{center}
\includegraphics[width=1.5in,height=3.0in]{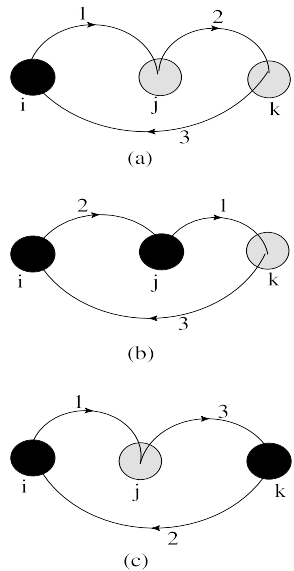}
\caption{Closed-loop hopping processes contributing to effective interaction term $V_n^i$ in
third-order 
perturbation theory.
Here filled (empty) circles 
correspond to sites with (without) hard-core-bosons. The numbers 1, 2, and 3 represent hopping 
sequence.}
\label{fey2}
\end{center}
\end{figure}
\begin{figure}[]
\begin{center}
\includegraphics[width=1.3in,height=3.0in]{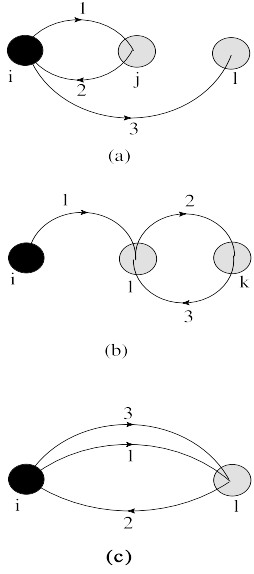}
\caption{Hopping processes (involving closed loops) contributing to effective hopping term 
$T_{Cn}^{li}$ in
third-order 
perturbation theory. Filled (empty) circles represent occupied (unoccupied) sites.
}
\label{fey3}
\end{center}
\end{figure}

In this appendix, we will show that the third-order
perturbation theory also produces a term that has the same
eigenstates as IRHM. To this end, we obtain the following third-order
 perturbation term in the effective Hamiltonian:
\begin{eqnarray}
\!\!\!H^{(3)}\!=\sum_{m\neq 0,n \neq 0}\!\!\!\! \frac{{_{ph}}\!\langle0|H_{I}|m\rangle_{ph}
~{_{ph}}\!\langle m|H_{I}|n\rangle_{ph}
~{_{ph}}\!\langle n|H_{I}|0\rangle_{ph}}
{{\Delta E_m^{ph}}{\Delta E_n^{ph}}}  . 
\nonumber \\ 
\label{H3}
\end{eqnarray}
Here $\Delta E_{m}^{ph}= \omega_{m}-\omega_{0}$.
Evaluation of $H^{(3)}$ leads to various hopping terms and interaction terms.
\begin{eqnarray}
H^{(3)}=\sum_{i,l\neq i} \left [\sum_{n=1}^{6} t_n T_n^{li} 
   + \sum_{n=1}^3 t_{cn} T_{Cn}^{li} \right ] + \sum_i \sum_{n=1}^3 v_n V^i_n , 
\nonumber \\ 
\label{tvtc}
\end{eqnarray}
where $t_n \sim (J^3 e^{-g^2})/(g^2 \omega)^2$,
 $t_{cn} \sim J^3 e^{-g^2}/(g\omega)^2$, and  $v_n \sim J^3 /(g^2 \omega)^2$
(as will be explained later).
We will demonstrate below that  $H^{(3)}$ is of the following form
\begin{eqnarray}
 H^{(3)} =\sum_{i,l > i} \left [ T(\sum_k n_k) b^{\dagger}_l b_i + {\rm H.c.} \right ] +  \sum_{i} 
V(\sum_k n_k) n_i  , 
\nonumber \\ 
\label{H3_form}
\end{eqnarray}
where $T$ and $V$ are functions of the total number operator $\sum_k n_k$.
 Since the IRHM commutes with the total number operator, $H^{(3)}$ has the same eigenstates as 
IRHM! 

 There are six open-loop hopping processes $T_n^{li}$ depicted in {figure} \ref{fey1}.  We analyze
them sequentially below.
\begin{eqnarray}
\!\!\!\!\!\!\!\!\!\!\!\!\!\!\!\!T_1^{li}
 &=& \sum_{k \neq i,l,j}\sum_{j \neq i,l} b^{\dagger}_l b_k b^{\dagger}_k b_j b^{\dagger}_j b_i 
\nonumber \\
 &=& \sum_{k \neq i,l,j} (1-b^{\dagger}_k b_k) \sum_{j \neq i, l} (1-b^{\dagger}_j b_j) 
b^{\dagger}_l b_i 
\nonumber \\
&=& \left [\sum_{k \neq i,l} (1-b^{\dagger}_k b_k) -1 \right ] \left [ \sum_{j \neq i, l} 
(1-b^{\dagger}_j b_j)\right ] b^{\dagger}_l b_i 
\nonumber \\
&=&\left [ \sum_{k \neq i,l} (1-b^{\dagger}_k b_k)-1 \right ] \left [ (N-2) - 
\sum_{j \neq l} b^{\dagger}_j b_j \right ] b^{\dagger}_l b_i
\nonumber \\
&=& \left [\sum_{k \neq i,l} (1-b^{\dagger}_k b_k)-1 \right ] \left [(N-1) - \sum_{j } b^{\dagger}_j 
b_j \right ] b^{\dagger}_l b_i
\nonumber \\
&=& \left [(N-1) - \sum_{j } b^{\dagger}_j b_j \right ] 
\left [\sum_{k \neq i,l} (1-b^{\dagger}_k b_k)-1 \right ]b^{\dagger}_l b_i
\nonumber \\
&=&\left [(N-1) - \sum_{j } b^{\dagger}_j b_j \right ] 
\left [(N-2) -\sum_{k } b^{\dagger}_k b_k \right ]b^{\dagger}_l b_i .
\nonumber \\
\end{eqnarray}
The second hopping
process $T_2^{li}$ in {figure} \ref{fey1} (b)  is given by 
\begin{eqnarray}
\!\!\!\!\!\!\!\!\!\!\!\!\!\!\!\!T_2^{li}
 &=& \sum_{k \neq i,l,j}\sum_{j \neq i,l} b^{\dagger}_j b_i b^{\dagger}_l b_k b^{\dagger}_k b_j 
\nonumber \\
 &=& \sum_{k \neq i,l,j} (1-b^{\dagger}_k b_k) \sum_{j \neq i, l} b^{\dagger}_j b_j b^{\dagger}_l 
b_i 
\nonumber \\
&=& \sum_{k \neq i,l} (1-b^{\dagger}_k b_k) \sum_{j \neq i, l} 
b^{\dagger}_j b_j b^{\dagger}_l b_i 
\nonumber \\
&=&\sum_{k \neq i,l} (1-b^{\dagger}_k b_k) \left [ 
\sum_{j } b^{\dagger}_j b_j -1 \right ] b^{\dagger}_l b_i
\nonumber \\
&=& \left [ 
\sum_{j } b^{\dagger}_j b_j -1 \right ]\left [ (N-1)-\sum_{k}b^{\dagger}_k b_k) \right ] 
b^{\dagger}_l b_i .
\end{eqnarray}
The hopping process $T_3^{li}$ in {figure} \ref{fey1} (c)  
is expressed as $T_3 ^{li}=\sum_{k \neq i,l,j}\sum_{j \neq i,l} b^{\dagger}_l b_k b^{\dagger}_j 
b_ib^{\dagger}_k b_j
 =T_2^{li}$.
The fourth hopping process $T_4^{li}$  in {figure} \ref{fey1} (d) is obtained as follows.
\begin{eqnarray}
\!\!\!\!\!\!\!\!\!\!\!\!\!\!\!\!T_4^{li}
 &=& \sum_{j \neq i,l, k}\sum_{k \neq i,l} b^{\dagger}_k b_j b^{\dagger}_j b_i b^{\dagger}_l b_k 
\nonumber \\
 &=& \sum_{j \neq i,l,k} (1-b^{\dagger}_j b_j) \sum_{k \neq i, l} b^{\dagger}_k b_k b^{\dagger}_l 
b_i 
\nonumber \\
&=& T_2^{li}  .
\end{eqnarray}
The hopping process $T_5^{li}$  in {figure} \ref{fey1} (e) yields
 $T_5^{li} =\sum_{j \neq i,l, k}\sum_{k \neq i,l} b^{\dagger}_k b_j  b^{\dagger}_l b_k b^{\dagger}_j 
b_i 
=T_4^{li}$.
 We  analyze below the last hopping 
process $T_6^{li}$ in {figure} \ref{fey1} (f).
\begin{eqnarray}
\!\!\!\!\!\!\!\!\!\!\!\!\!\!\!\!T_6^{li}
 &=& \sum_{k \neq i,l,j}\sum_{j \neq i,l} b^{\dagger}_j b_i b^{\dagger}_k b_j b^{\dagger}_l b_k 
\nonumber \\
 &=& \sum_{k \neq i,l,j} b^{\dagger}_k b_k \sum_{j \neq i, l} b^{\dagger}_j b_j b^{\dagger}_l b_i 
\nonumber \\
&=& \left [\sum_{k \neq i,l} b^{\dagger}_k b_k - 1 \right ] \sum_{j \neq i, l} 
b^{\dagger}_j b_j b^{\dagger}_l b_i 
\nonumber \\
&=&\left [\sum_{k \neq i,l} b^{\dagger}_k b_k - 1 \right ]\left [ 
\sum_{j } b^{\dagger}_j b_j -1 \right ] b^{\dagger}_l b_i
\nonumber \\
&=& \left [ 
\sum_{j } b^{\dagger}_j b_j -1 \right ]\left [ \sum_{k}b^{\dagger}_k b_k-2 \right ] b^{\dagger}_l 
b_i  .
\end{eqnarray}

We will now deal with closed-loop hopping processes such as those in {figure} \ref{fey2}.
These lead to effective interactions. The process $V_1^i$ in {figure} \ref{fey2} (a),  obtained from
figure \ref{fey1} (a) by setting $l=i$, is given as follows. 
\begin{eqnarray}
\!\!\!\!\!\!\!\!\!\!\!\!\!\!\!\!V_1^i
 &=& \sum_{k \neq i,j}\sum_{j \neq i} b^{\dagger}_i b_k b^{\dagger}_k b_j b^{\dagger}_j b_i 
\nonumber \\
 &=& \sum_{k \neq i,j} (1-b^{\dagger}_k b_k) \sum_{j \neq i} (1-b^{\dagger}_j b_j) b^{\dagger}_i 
b_i 
\nonumber \\
&=& \left [\sum_{k \neq i} (1-b^{\dagger}_k b_k) -1 \right ] \left [ \sum_{j \neq i} 
(1-b^{\dagger}_j b_j)\right ] b^{\dagger}_i b_i 
\nonumber \\
&=&\left [(N) - \sum_{j } b^{\dagger}_j b_j \right ] 
\left [(N-1) -\sum_{k } b^{\dagger}_k b_k \right ]b^{\dagger}_i b_i .
\end{eqnarray}
Next, the hopping process $V_2^i$ corresponding to closed loop in {figure} \ref{fey2} (b)
 is obtained from {figure} \ref{fey1} (c) by taking $l=i$.
\begin{eqnarray}
\!\!\!\!\!\!\!\!\!\!\!\!\!\!\!\!V_2^i
 &=& \sum_{k \neq i,j}\sum_{j \neq i} b^{\dagger}_i b_k b^{\dagger}_j b_i b^{\dagger}_k b_j 
\nonumber \\
 &=& \sum_{k \neq i,j} (1-b^{\dagger}_k b_k) \sum_{j \neq i} b^{\dagger}_j b_j b^{\dagger}_i b_i 
\nonumber \\
&=& \sum_{k \neq i} (1-b^{\dagger}_k b_k) \sum_{j \neq i} 
b^{\dagger}_j b_j b^{\dagger}_i b_i 
\nonumber \\
&=&\sum_{k \neq i} (1-b^{\dagger}_k b_k) \left [ 
\sum_{j } b^{\dagger}_j b_j -1 \right ] b^{\dagger}_i b_i
\nonumber \\
&=& \left [ 
\sum_{j } b^{\dagger}_j b_j -1 \right ]\left [ (N)-\sum_{k}b^{\dagger}_k b_k) \right ] b^{\dagger}_i 
b_i .
\end{eqnarray}
Lastly, the hopping $V_3^i$ [depicted by the closed loop in {figure} \ref{fey2} (c)]
 is obtained from {figure} \ref{fey1} (e) by setting $l=i$.
\begin{eqnarray}
\!\!\!\!\!\!\!\!\!\!\!\!\!\!\!\!V_3^i
 &=& \sum_{j \neq i, k}\sum_{k \neq i} b^{\dagger}_k b_j b^{\dagger}_i b_k b^{\dagger}_j b_i
\nonumber \\
 &=& \sum_{j \neq i,k} (1-b^{\dagger}_j b_j) \sum_{k \neq i} b^{\dagger}_k b_k b^{\dagger}_i b_i 
\nonumber \\
&=& V_2^i  .
\end{eqnarray}

Finally, we consider {figures} \ref{fey3} (a), (b), and (c) 
 which deal with effective hopping terms $T_{Cn}^{li}$ involving closed loops. The effective hopping
 term $T_{C1}^{li}$,
corresponding to {figure} \ref{fey3} (a), is obtained by setting $k=i$ in {figure} \ref{fey1} (a):
\begin{eqnarray}
\!\!\!\!\!\!\!\!\!\!\!\!\!\!\!\!T_{C1}^{li}
 &=& \sum_{j \neq i, l} b^{\dagger}_l b_i b^{\dagger}_i b_j b^{\dagger}_j b_i 
\nonumber \\
 &=&  \sum_{j \neq i, l} (1-b^{\dagger}_j b_j) b^{\dagger}_l b_i 
\nonumber \\
&=& \left [ (N-2) - 
\sum_{j \neq l} b^{\dagger}_j b_j \right ] b^{\dagger}_l b_i
\nonumber \\
&=&  \left [(N-1) - \sum_{j } b^{\dagger}_j b_j \right ] b^{\dagger}_l b_i .
\end{eqnarray}
To obtain the effective hopping term $T_{C2}^{li}$
corresponding to {figure} \ref{fey3} (b), we take $j=l$ in {figure} \ref{fey1} (a):
\begin{eqnarray}
\!\!\!\!\!\!\!\!\!\!\!\!\!\!\!\!T_{C2}^{li}
 &=& \sum_{k \neq i,l}
b^{\dagger}_l b_k b^{\dagger}_k b_l b^{\dagger}_l b_i 
\nonumber \\
 &=& \sum_{k \neq i,l} (1-b^{\dagger}_k b_k)  b^{\dagger}_l b_i 
\nonumber \\
&=& \left [ (N-2) - 
\sum_{k \neq l} b^{\dagger}_k b_k \right ] b^{\dagger}_l b_i
\nonumber \\
&=& \left [(N-1) - \sum_{k } b^{\dagger}_k b_k \right ] b^{\dagger}_l b_i 
\nonumber \\
&=& T_{C1}^{li} .
\end{eqnarray}
The effective hopping term $T_{C3}^{li}$ depicted in {figure} \ref{fey3} (c) 
[upon setting $k=i$ and $j=l$ in {figure} \ref{fey1} (a)] is given by
\begin{eqnarray}
\!\!\!\!\!\!\!\!\!\!\!\!\!\!\!\!T_{C3}^{li}
 = b^{\dagger}_l b_i b^{\dagger}_i b_l b^{\dagger}_l b_i 
= b^{\dagger}_l b_i .
\end{eqnarray}

 Thus we have shown that $H^{(3)} $ contains effective hopping terms 
{$\sum_{i,l>i} [T(\sum_k n_k) b^{\dagger}_l b_i$ \\ $+ {\rm H.c.]} $}
 and effective interaction terms ($\sum_{i} V(\sum_k n_k) n_i$). Since 
 $T$ and $V$ are functions of the total number operator,
 $H^{(3)}$ and IRHM have the same eigenstates. These arguments can be extended to even
higher-order perturbation theory to show that the effective Hamiltonian (after taking all
orders of perturbation into account) will give the same eigenstates as IRHM!

\begin{figure}[]
\begin{center}
\includegraphics[width=1.5in,height=3.0in]{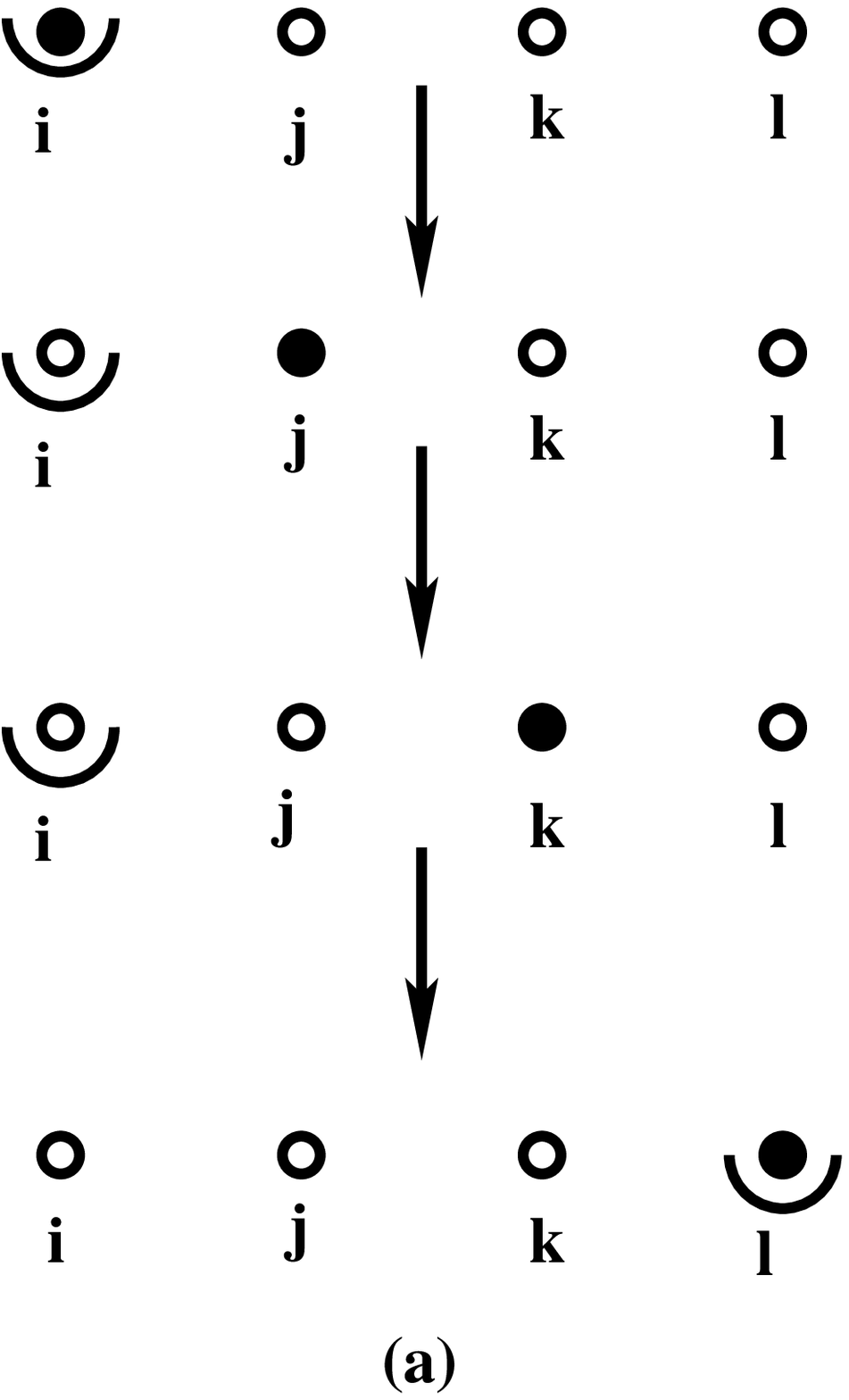}\\
\vspace{.3cm}
\includegraphics[width=2.5in,height=2.0in]{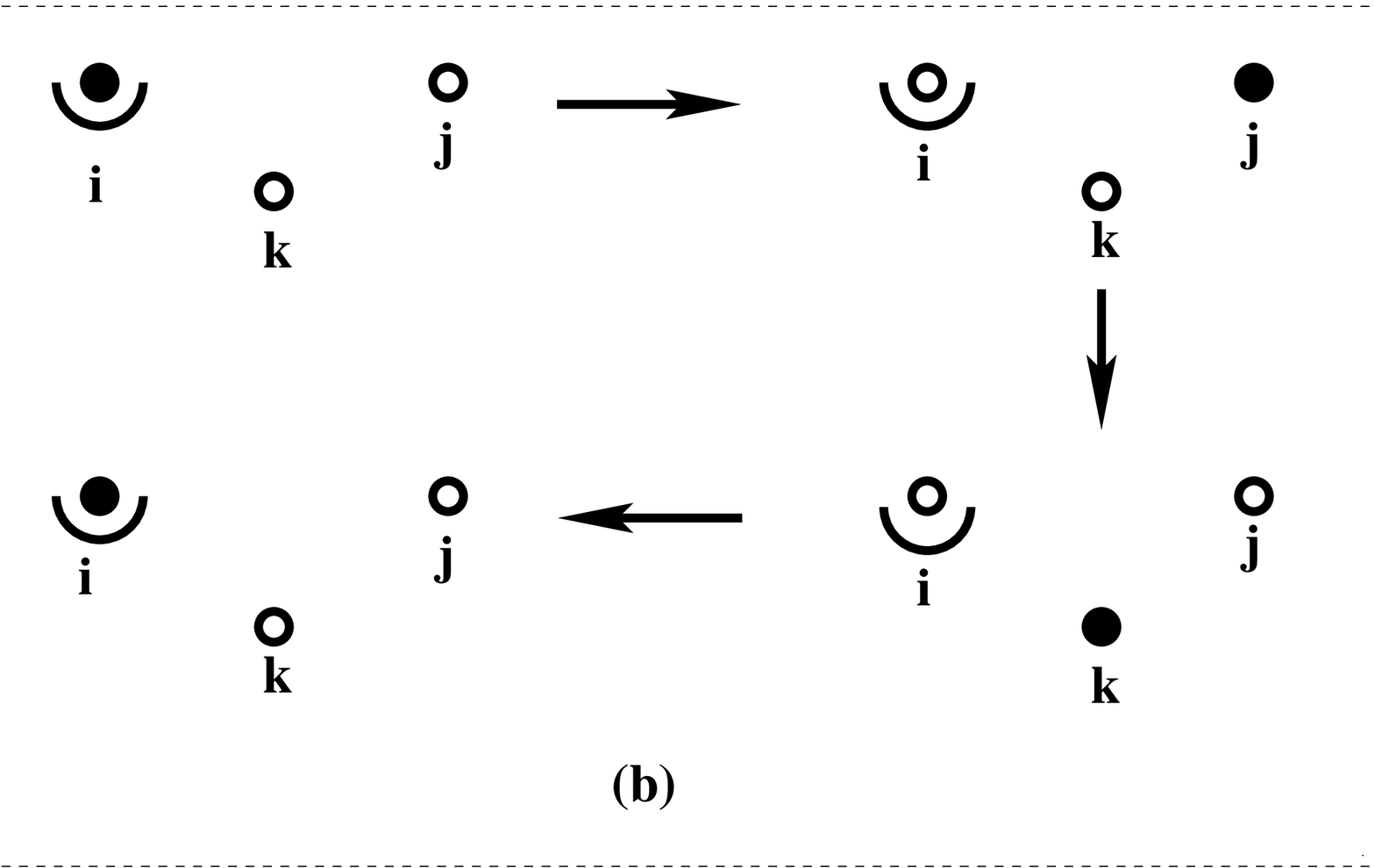}\\
\vspace{.3cm}
\includegraphics[width=2.5in,height=2.0in]{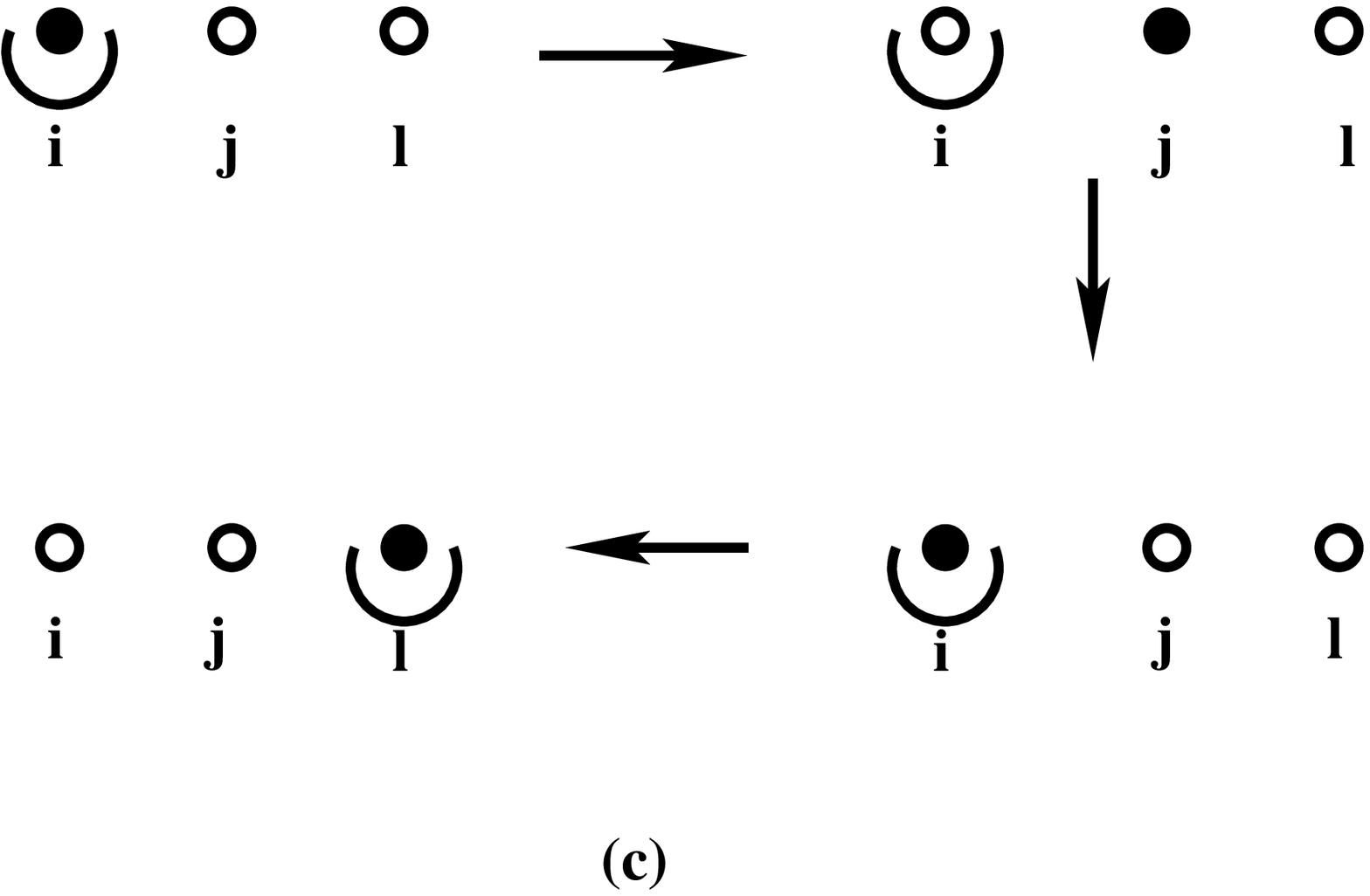}
\caption{Schematic diagrams (a), (b), and (c), corresponding
to the hopping processes depicted in {figure} \ref{fey1} (a), {figure} \ref{fey2} (a), 
and {figure} \ref{fey3} (a), respectively,  yield coefficients
$t_n$, $v_n$, and $t_{cn}$, respectively.  The intermediate states give the typical dominant
contributions. Here empty circles correspond to empty sites, while filled circles
 indicate particle positions. Parabolic curve
at a site depicts full distortion at that site with
corresponding energy $-g^2 \omega$ ($+g^2 \omega$) if the hard-core-boson is present
(absent) at that site.
}
\label{schemat}
\end{center}
\end{figure}
We will now explain the expressions for the coefficients $t_n$, $v_n$, and $t_{cn}$
in equation (\ref{tvtc}), 
 obtained from third-order
perturbation theory, using typical schematic
diagrams shown in {figure} \ref{schemat} [for details of corresponding
diagrams and analysis in second order perturbation, see reference \cite{srsypbl}]. We consider two 
distinct time scales
associated with  hopping processes between two sites:
(i) $\sim 1/(Je^{-g^2})$ corresponding to either full distortion at a site
to form a small polaronic potential well (of energy $-g^2 \omega$) or
full relaxation from the small polaronic distortion and (ii) $\sim 1/J$
related to negligible distortion/relaxation at a site. 
The coefficient $t_n$ corresponds
to the typical dominant distortion processes shown schematically in {figure} \ref{schemat} (a)
with the pertinent typical  hopping processes being depicted in
{figure} \ref{fey1} (a). In {figure} \ref{schemat} (a), after the HCB hops
 away from the initial site, the intermediate states
have the same distortion as the initial state. Next,
when the HCB hops to its final site there is a distortion
at this final site with a concomitant relaxation at the initial site.
Hence the contribution to the coefficient $t_n$ becomes
$J/(2 g^2 \omega) \times J/(2 g^2 \omega) \times J e^{-g^2} \sim J^3 e^{-g^2}/(g^2 \omega)^2$.
As regards coefficient $v_n$, it can be deduced based on the typical dominant 
hopping-cum-distortion 
processes depicted in {figure} \ref{schemat} (b) which typifies the hopping processes in {figure} 
\ref{fey2} (a).
In {figure} \ref{schemat} (b), when the particle hops to different sites and reaches finally the 
initial site,
there is no change in distortion at any site. Hence $v_n$ can be estimated to be
$J/(2 g^2 \omega) \times J/(2 g^2 \omega) \times J  \sim J^3 /(g^2 \omega)^2$.
Lastly, we obtain the coefficient $t_{cn}$ by considering the 
typical dominant diagram in {figure} \ref{schemat} (c)
corresponding to the typical process in {figure} \ref{fey3} (a). In {figure} \ref{schemat} (c),
where the first intermediate state depicts the particle hopping but leaving the distortion 
unchanged,
we get a contribution $J/(2g^2 \omega)$; for the next intermediate state, where the HCB returns to
the initial site, the initial site has to undergo a slight relaxation (involving absorbing a phonon
so as to yield a non-zero denominator in the perturbation theory) leading to the contribution 
$J/\omega$;
and lastly, when the HCB hops to the final site, there is a distortion
at the final site with a simultaneous relaxation at the initial site thereby producing
a contribution $J e^{-g^2}$. Thus we calculate $t_{cn}$ to be
 $J/(2g^2 \omega) \times J/\omega \times J e^{-g^2} \sim J^3 e^{-g^2}/(g \omega)^2$.

 \section{}

 In this appendix we will evaluate the various terms in  master equation (\ref{mark}). 
Defining $\{|n\rangle_{ph}\}$ as the basis set for phonons, therefore, we can write
the master equation (equation (\ref{mark})) as:

\begin{widetext}

{\begin{eqnarray}
\!\!\!\!\!\!\!\!\!\!\!
\frac{d \tilde{\rho}_s(t)}{dt} &=&-i  \sum_{n}~ _{ph}\langle n| [\tilde{H}_I(t), \rho_s(0) \otimes 
R_o] |n \rangle_{ph}
\nonumber \\
&& - \sum_n \int_0^\infty d\tau \left[~ _{ph}\langle n |
 \tilde{H}_I(t)\tilde{H}_I(t-\tau)\tilde{\rho}_s(t) \otimes R_o|n\rangle_{ph}  \right .
 \nonumber \\
&& ~~~~~~~~~~~~~~~~ - ~_{ph}\langle n| \tilde{H}_I(t)\tilde{\rho}_s(t) \otimes R_o 
\tilde{H}_I(t-\tau) |n\rangle_{ph} \nonumber \\
&& ~~~~~~~~~~~~~~~~ - ~_{ph}\langle n| \tilde{H}_I(t-\tau)\tilde{\rho}_s(t) \otimes R_o 
\tilde{H}_I(t) |n\rangle_{ph}
 \nonumber \\
&& ~~~~~~~~~~~~~~~~
 + \left .
~_{ph}\langle n | \tilde{\rho}_s(t) \otimes R_o \tilde{H}_I(t-\tau)\tilde{H}_I(t)|n\rangle_{ph}
 \right] .
\label{mas}
\end{eqnarray}}

In order to simplify the above master equation, we need to evaluate  
the time evolution of the operators involved in $H_I$. Considering the 
second term in the equation (\ref{mas}), yields
{ \small
\begin{eqnarray}
_{ph}\langle n |
 \tilde{H}_I(t)\tilde{H}_I(t-\tau)\tilde{\rho}_s(t) \otimes R_o|n\rangle_{ph}
  =\sum_m e^{iH_s t} {_{ph}\langle n | H_I |m\rangle_{ph}} e^{-iH_s t} ~
    e^{iH_s (t-\tau)}  {_{ph}\langle m| H_I|n\rangle_{ph}} e^{-iH_s( t-\tau)} \tilde{\rho}_s(t)
\frac{e^{-\beta \omega_n}}{Z} e^{i(\omega_n-\omega_m)\tau} .\nonumber \\
\label{time}
\end{eqnarray}}

In momentum space, we express HCB operators as:
$ b^{\dagger}_j =
 \frac{1}{\sqrt{N}}\sum_k e^{ikr_j }~ b^{\dagger}_k$ and $ b_j = \frac{1}{\sqrt{N}}\sum_k e^{-ikr_j 
}~ b_k$; 
then, it is important to note that the hopping term 
in the system Hamiltonian can be written as:
\begin{eqnarray}
0.5 J \sum_{i,j>i}( e^{-g^2} b^{\dagger}_i  b_{j} +{\rm H.c.}) &=& 
0.5 J e^{-g^2} \left[ \sum_{i, j}b^{\dagger}_i  b_{j} - \sum_{i}b^{\dagger}_i  b_{i} \right] 
\nonumber \\
&=& 0.5 J^{\star}(\frac{N}{N-1}) e^{-g^2} \hat{n}_0 - 0.5 J e^{-g^2} \hat{N_p}  \nonumber \\
&=&\sum_{k} \epsilon_k b^{\dagger}_k b_k , 
\end{eqnarray}
where we used $ J = J^{\star}/(N-1)$,  $\hat{N_p}\equiv \sum_k b^{\dagger}_k b_k$ and 
 $ \hat{n}_0 \equiv b^{\dagger}_0 b_0$ (i.e., 
the particle number in momentum $k=0$ state). 
{Here it should be mentioned that using HCBs instead of spins has enabled us to
obtain (with ease) the excitation spectrum $\epsilon_k$ which is crucial
for the analysis given below}.
Let $\{ |q \rangle_s\}$  denote the  complete set of 
energy eigenstates (with eigenenergies $E_q^s$) of the system Hamiltonian $H_s$;
 then we can write:
{\begin{eqnarray}
 e^{iH_s t} H_I  e^{-iH_s t} 
= 0.5 J e^{-g^2}\sum_{l,j>l}\sum_{q, q^{\prime}} |q \rangle_s {_s \!\langle} q| 
 e^{iH_st} \left[  \frac{1}{N}  
\sum_{k, p}b^{\dagger}_k b_p e^{i(k r_l - p r_j)} \right] e^{-iH_st} |q^{\prime} \rangle_s 
{_s\!\langle} q^{\prime} |  
  \{\mathcal S^{{lj}^\dagger}_+ \mathcal S^{lj}_{-}-1\}  \nonumber \\
 ~~~~~~~~~~~~~~~~~~~~~~~~~~~~~~~~~~~~~~~~~~~~~~~~~~~~~~~~+ {\rm H. c.} ,
  \nonumber \\
\end{eqnarray}}
which implies
\begin{eqnarray}
 e^{iH_s t}~ _{ph}\langle n | H_I |m\rangle_{ph} e^{-iH_s t}
 = 
 \sum_{q, q^{\prime}} |q \rangle_s {_s\!\langle} q| ~ _{ph}\langle n | H_I |m\rangle_{ph} 
 |q^{\prime} \rangle_s {_s\!\langle} q^{\prime} | e^{i(E_q^s-E_{q^{\prime}}^s)t} ,
 \nonumber \\
\label{TE}
\end{eqnarray}
where $|E_q^s-E_{q^{\prime}}^s| = 0.5 J^{\star} (\frac{N}{N-1}) e^{-g^2}$ or $0$ . {Here we have 
taken  
 the total number of HCBs to be conserved; then,
 only the hopping term in $H_s$ will contribute to the particle excitation energy.}
Substituting equation (\ref{TE}) in equation (\ref{time}), we get
{\begin{eqnarray}
 _{ph}\langle n |
 \tilde{H}_I(t)\tilde{H}_I(t-\tau)\tilde{\rho}_s(t) \otimes R_o|n\rangle_{ph}
 =
 \sum_m \sum_{q, q^{\prime}, q^{\prime \prime}} \left[ \{|q \rangle_s {_s\!\langle} q| ~ 
_{ph}\langle 
n | H_I |m\rangle_{ph} 
|q^{\prime} \rangle_s {_s\!\langle} q^{\prime} |~ _{ph}\langle m | H_I |n\rangle_{ph}
 |q^{\prime \prime} \rangle_s {_s\!\langle} q^{\prime\prime} | \}
 \right.
\nonumber \\
~~~~~~~\left.  \times
e^{i[(E_q^s-E_{q^{\prime}}^s)t+(E_{q^{\prime}}^s-E_{q^{\prime\prime}}^s)(t-\tau)]}   
  \tilde{\rho}_s(t)\frac{e^{-\beta \omega_n}}{Z} 
e^{i(\omega_n-\omega_m)\tau} \right]. 
\label{time1}
\end{eqnarray}}
 Thus under the assumption of $J^{\star} e^{-g^2} < < \omega$,
it follows that $|\omega_n - \omega_m| > > |E_q^s - E_{q^{\prime}}^s|$ and
$|\omega_n - \omega_m| > > |E_{q^{\prime}}^s - E_{q^{\prime \prime}}^s|$;
hence in equation (\ref{time1}),  we can 
take 
$e^{i[(E_q^s-E_{q^{\prime}}^s)t]} =1$
 and
$e^{i[(E_{q^{\prime}}^s-E_{q^{\prime\prime}}^s)(t-\tau)]} =1$
 which implies that 
{
we do not get terms
producing decay}. The resultant equation is
\begin{eqnarray}
\!\!\!\!\!\!\!
 _{ph}\langle n |
 \tilde{H}_I(t)\tilde{H}_I(t-\tau)\tilde{\rho}_s(t) \otimes R_o|n\rangle_{ph} =
\sum_m {_{ph}\langle n | H_I |m\rangle_{ph}} ~
     {_{ph}\langle m| H_I|n\rangle_{ph}} ~ \tilde{\rho}_s(t) 
 \frac{e^{-\beta \omega_n}}{Z} e^{i(\omega_n-\omega_m)\tau} .
 \nonumber \\
\label{time2}
\end{eqnarray}

Carrying out the same analysis on the remaining (i.e., third, fourth, and fifth) terms in the 
master equation, we write equation (\ref{mas})
as:

{\begin{eqnarray}
~~~~~\frac{d \tilde{\rho}_s(t)}{dt} &=& -i  \sum_{n}~ _{ph}\langle n| 
[\tilde{H}_I(t), 
\tilde{\rho}_s(0) \otimes R_o] |n \rangle_{ph}
\nonumber \\
&& - \sum_{n,m} \int_0^\infty d\tau \left[ _{ph}\langle n | H_I |m\rangle_{ph}~
_{ph}\langle m| H_I|n\rangle_{ph} ~ \tilde{\rho}_s(t) 
 \frac{e^{-\beta \omega_n}}{Z} e^{i(\omega_n-\omega_m)\tau} \right.
 \nonumber \\
&& ~~~~~~~~~~~~~~~~ - ~_{ph}\langle n | H_I |m\rangle_{ph} ~ \tilde{\rho}_s(t) ~ _{ph}\langle m|
 H_I|n\rangle_{ph} \frac{e^{-\beta \omega_m}}{Z} e^{i(\omega_n-\omega_m)\tau}
\nonumber \\ 
&&~~~~~~~~~~~~~~~~ - ~ _{ph}\langle n | H_I |m\rangle_{ph} ~ \tilde{\rho}_s(t) ~  _{ph}\langle m| 
H_I|n\rangle_{ph} 
\frac{e^{-\beta \omega_m}}{Z} e^{-i(\omega_n-\omega_m)\tau} 
 \nonumber \\
&& ~~~~~~~~~~~~~~~~ + \left. \tilde{\rho}_s(t)~ _{ph}\langle n | H_I |m\rangle_{ph}~_{ph}\langle m| 
H_I|n\rangle_{ph} 
  \frac{e^{-\beta \omega_n}}{Z} e^{-i(\omega_n-\omega_m)\tau}
\right] .
\end{eqnarray}}
Next, we evaluate the first term in the above equation and show that it is zero at 
$T=0$. We observe that
{\begin{eqnarray}
 Tr_R[\tilde{H}_I(t)R_o]&=& \sum_{n}~ _{ph}\langle n |\tilde{H}_I(t)R_o|n \rangle_{ph} \nonumber \\
&=& 0.5 J  e^{-g^2}\sum_{l,j\neq l} \left[ e^{iH_s t} b^{\dagger}_l b_j e^{-iH_st}~ 
_{ph}\langle 0 |  \{\mathcal S^{{lj}^\dagger}_+ \mathcal S^{lj}_{-} 
 -1\}  |0 \rangle_{ph} \right] \nonumber \\ &=& 0 .
\end{eqnarray}}
Thus, we have $  \sum_{n}~ _{ph}\langle n| [\tilde{H}_I(t), \rho_s(0) \otimes R_o] |n 
\rangle_{ph}=0$
and the master equation at $T=0$ simplifies as in equation (\ref{21}).

\end{widetext}


\begin{thebibliography}{47}
\bibitem{indexk}
The bath at each site is $\omega \sum_{k,j} a^{\dagger}_{k,j} a_{k,j}$ and $a_{k,j}$
is the phonon destruction operator. Here,
we suppressed the index $k$ (which represents the phonon modes) as it does not change
any results.
\bibitem{gl} If the perturbation $(H_s-H_{\rm IRHM})$ is adiabatically turned on ( i.e.,
 $H_s(t)=H_{\rm IRHM}+(H_s-H_{\rm IRHM})e^{-\lambda| t |}$, 
where $\lambda$ is a small positive quantity), the eigenstate $| \psi\rangle$ of our starting 
Hamiltonian $H_s$ can be unitarily connected to 
the eigenstate $| \phi\rangle$ of $H_{\rm IRHM}$,
 i.e., $| \psi(0)\rangle=U_I^{\dag}(-\infty)| \phi\rangle$, 
where $U_I(t)=e^{-i\int\limits_0^t(H_s-H_{\rm IRHM})e^{-\lambda | t' |}dt'}$.
\end{thebibliography}
\end{document}